\documentclass[a4paper,showpacs]{revtex4}
\usepackage{tabularx}
\usepackage{amsmath}
\usepackage{amssymb}
\usepackage{graphicx}
\usepackage{dcolumn}
\usepackage{bm}
\usepackage{graphicx}
\usepackage{epstopdf}

\begin{document}

\title{Dzyaloshinskii-Moriya Interaction and Spiral Order in Spin-orbit
Coupled Optical Lattices}
\author{Ming Gong$^{1,2}$}
\author{Yinyin Qian$^{1}$}
\author{Mi Yan$^{3}$}
\author{V. W. Scarola$^{3}$}
\thanks{Corresponding author, email: scarola@vt.edu}
\author{Chuanwei Zhang$^{1}$}
\thanks{Corresponding author, email: chuanwei.zhang@utdallas.edu}

\begin{abstract}
We show that the recent experimental realization of spin-orbit coupling in
ultracold atomic gases can be used to study different types of spin spiral order
and resulting multiferroic effects. Spin-orbit coupling in optical lattices
can give rise to the Dzyaloshinskii-Moriya (DM) spin interaction which is
essential for spin spiral order. By taking into account spin-orbit coupling
and an external Zeeman field, we derive an effective spin model in the Mott
insulator regime at half filling and demonstrate that the DM interaction in
optical lattices can be made extremely strong with realistic experimental
parameters. The rich finite temperature phase diagrams of the effective spin
models for fermions and bosons are obtained via classical Monte Carlo
simulations.
\end{abstract}

\affiliation{$^{1}$Department of Physics, the University of Texas at Dallas, Richardson, Texas, 75080 USA \\
$^{2}$ Department of Physics and Center for Quantum Coherence, The Chinese University of Hong Kong, Shatin, N.T., Hong Kong, China \\
$^{3}$Department of Physics, Virginia Tech, Blacksburg, Virginia 24061 USA}
\pacs{67.85.Hj, 03.75.Lm, 67.85.Fg}
\maketitle
\section*{\bf Introduction}
\label{intro}
The interplay between ferroelectric and ferromagnetic order in complex
multiferroic materials presents a set of compelling fundamental condensed
matter physics problems with potential multifunctional device applications
\cite{reviews-1,reviews-2,reviews-3,reviews-4}. Ferroelectric and ferromagnetic order compete and normally
cannot exist simultaneously in conventional materials. While in some
strongly correlated materials, such as the perovskite transition metal
oxides \cite{MF-1, MF-2,MF-3,MF-4,MF-5,MF-6}, these two phenomena can occur
simultaneously due to strong correlation. Nowadays construction and design
of high-$T_{c}$ magnetic ferroelectrics is still an open and active area of
research \cite{Jing}. These materials incorporate different types of
interactions, including electron-electron interactions, electron-phonon
interactions, spin-orbit (SO) couplings, lattice defects, and disorder,
making the determination of multiferroic mechanisms a remarkable challenge
for most materials \cite{Sergienko,Katsura}. In this context an unbiased and
direct method to explore multiferroic behavior in an ideal setting is highly
appealing.

On the other hand, the realization of a superfluid to Mott insulator
transition of ultracold atoms in optical lattices \cite{Greiner} opens
fascinating prospects \cite{blochreview} for the emulation of a large
variety of novel magnetic states \cite{Kuklov,Duan,Altman} and other
strongly correlated phases found in solids because of the high
controllability and the lack of disorder in optical lattices. For instance,
it has been shown \cite{Kuklov,Duan} that the effective Hamiltonian of
spin-1/2 atoms in optical lattices is the XXZ Heisenberg model in the deep
Mott insulator regime. On the experimental side, superexchange interactions
between two neighboring sites have already been demonstrated \cite{Trotzky}
and quantum simulation of frustrated classical magnetism in triangular optical
lattices has also been realized \cite{Struck}. These experimental
achievements mark the first steps towards the quantum simulation of possible
magnetic phase transitions in optical lattices.

In this paper, we show that the power of optical lattice systems to emulate
magnetism can be combined with recent experimental developments \cite%
{NIST,Pan,Zhangjing1,Zhangjing2} realizing SO coupling to emulate
multiferroic behavior. Recently, SO coupled optical lattices have been
realized in experiments for both bosons \cite{Ian} and fermions \cite%
{Zwierwei}, where interesting phenomena such as flat bands \cite%
{Zwierwei,Yongping1,Lin1} can be observed. The main findings of this work are
the following: (I) We incorporate spin-orbit and Zeeman
coupling into an effective Hamiltonian for spin-1/2 fermions and bosons
in optical lattices in the large interaction limit. We show that SO coupling
leads to an effective in-plane Dzyaloshinskii-Moriya (DM) term, an essential
ingredient in models of spiral order and multiferroic effects in general.
The DM term is of the same order as the Heisenberg coupling constant. (II)
We study the finite temperature phase diagram of the effective spin model 
using classical Monte Carlo (MC). We find that competing types of spiral
order depend strongly on both SO and effective Zeeman coupling strength. (III) 
We find that the critical temperature for the spiral order can
be of the same order as the Heisenberg coupling constant. Thus, if magnetic
quantum phase transitions can be emulated in optical lattices, then spiral
order and multiferroic-based models can also be realized in the same setup with the
inclusion of SO coupling.

\section*{\bf Results}
\label{results}

\noindent \textbf{{\small Effective Hamiltonian.}}
We consider spin-1/2 ultracold atoms loaded
into a two-dimensional (2D) square optical lattice. We restrict ourselves to
the deep Mott insulator regime where the charge/mass degree of freedom is
frozen while the spin degree of freedom remains active. Here the atomic
hyperfine levels map onto effective spin states. The scattering length
between atoms in optical lattices can be controlled by a Feshbach
resonance.  Certain atoms, e.g., $^{40}$K, exhibit considerable tunability\cite{Kohl}. To derive the inter-spin interaction in this regime
we first consider a two-site tight-binding model,
\begin{equation}
H=-\sum_{\sigma }t_{\sigma }c_{1\sigma }^{\dagger }c_{2\sigma }^{%
\phantom{\dagger}}+V_{\text{so}}+V_{z}+{\frac{1}{2}}\sum_{i,\sigma \sigma
^{\prime }}U_{\sigma \sigma ^{\prime }}:n_{i\sigma }n_{i\sigma ^{\prime }}:,
\label{eq-H}
\end{equation}%
where $c_{i\sigma }^{\dagger }$ creates a particle (either a boson or a
fermion) in a Wannier state, $w_{i,\sigma }$, localized at a site $i$ and in
a spin state $\sigma \in \{\uparrow ,\downarrow \}$. $n_{i\sigma
}=c_{i\sigma }^{\dagger }c_{i\sigma }^{\phantom{\dagger}}$ is the number operator. The tunneling and interaction matrix elements are $t_{\sigma}=\int d\mathbf{x}w_{i,\sigma}^{\ast
}[\mathbf{p}^{2}/2m+V(\mathbf{x})]w_{i+1,\sigma}$ and $U_{\sigma
\sigma ^{\prime }}=g_{\sigma \sigma ^{\prime }}\int d\mathbf{x}|w_{i,\sigma
}|^{2}|w_{i,\sigma ^{\prime }}|^{2}$, respectively, where $g_{\sigma \sigma ^{\prime }}$
is the interaction strength between species $\sigma $ and $\sigma ^{\prime }$, $m$ is the mass of the atom, 
and $V(\mathbf{x})$ is a lattice potential. Here $::$ denotes normal ordering. For a general theory the tunneling is
assumed to be spin dependent, which is a feature unique to ultracold atom
systems \cite{Duan, Altman}. The second term is the Rashba SO coupling \cite
{YPPRL},  written in the continuum as $\gamma (p_{x}\sigma _{y}-p_{y}\sigma _{x})$. But on a lattice it can be written as
\begin{equation}
V_{\text{so}}=i\lambda c_{i}^{\dagger }\mathbf{e}_{z}\cdot (\boldsymbol{\sigma} \times
\mathbf{d})c_{j}^{\phantom{\dagger}}+h.c.,
\label{eq-SOC}
\end{equation}%
where $c_{i}^{\dagger }=(c_{i\uparrow }^{\dagger },c_{i\downarrow }^{\dagger
})$, $\boldsymbol{\sigma}$ denotes Pauli matrices, and $\lambda=-i\gamma \int d\mathbf{x} w_{i}^{\ast
} p_x w_{i+\mathbf{e}_{x}} $ is the SO coupling strength. $\mathbf{d\equiv }\left( dx,dy\right) $ is the vector from a site at position $\boldsymbol{r}_{j}$ to a site at $\boldsymbol{r}_{i}$, where $dx=\left(\boldsymbol{r}_{i}-\boldsymbol{r}_{j}\right)\cdot \mathbf{e}_{x}$ and $dy=\left(\boldsymbol{r}_{i}-\boldsymbol{r}_{j}\right)\cdot \mathbf{e}_{y}$. Eq. \ref{eq-SOC}  describes the tunneling between neighboring sites paired with a spin flip. The magnitude and sign of $%
\lambda $ can be tuned in experiments using coherent destructive tunneling
methods \cite{YP}. The third term is the external Zeeman field $%
V_{z}=\sum_{i,\sigma }\Upsilon _{\sigma }n_{i\sigma }$ with $\Upsilon
_{\sigma }=\pm \Upsilon /2$.

In the deep Mott insulator regime, the degeneracy in spin configurations is lifted by second order virtual processes. The effective Hamiltonian $H_{\text{%
eff}}$ can be obtained
using perturbation theory. We take the Mott insulator as the unperturbed state and derive the corrections of the effective Hamiltonian by the standard Schrieffer-Wolf transformation   \cite{WS,Duan}. The Schrieffer-Wolf transformation applies a canonical transformation $H_{\text{%
eff}}=e^{iS}He^{-iS}$ to obtain the second order Hamiltonian $H_{\text{eff}%
}=H_{0}+{\frac{1}{2}}[iS,V]$ by eliminating the first order term
using $V=-[iS,H_{0}]$. In the spin representation we define $\mathbf{S}%
_{i}=\sum_{ss}c_{is}^{\dagger }\mathbf{\sigma }_{ss^{\prime }}c_{is^{\prime
}}^{\phantom{\dagger}}$, and extend the two-site model to the whole lattice,
yielding%
\begin{eqnarray}
H_{\text{eff}} &=&\sum_{\langle i,j\rangle }\sum_{\alpha =x,y,z}J_{\alpha
}S_{i}^{\alpha }S_{j}^{\alpha }+\sum_{i}\mathbf{B}\cdot \mathbf{S}_{i}+
\notag \\
&&\sum_{ij}\mathbf{D}_{ij}\cdot (\mathbf{S}_{i}\times \mathbf{S}_{j})+%
\mathbf{S}_{i}\cdot \Gamma _{ij}\cdot \mathbf{S}_{j}.  \label{eff-HF}
\end{eqnarray}%
The first two terms are Heisenberg exchange and Zeeman terms,
respectively, while the last two terms arise from SO coupling. In solid
state systems the third term is called the DM interaction \cite{DM-1, DM-2}, which
is believed to drive multiferroic behavior.  The definition of the $\bf{D}$ vector and the $\Gamma$ tensor 
will be presented below.  The structure of these terms can be derived from basic symmetry analyses but the coefficients must be computed microscopically. In the following we derive the
coefficients in Eq. \ref{eff-HF} by considering the coupling between four
internal degenerate ground states $|\alpha \rangle \in \left\{ |\uparrow
;\uparrow \rangle ,|\uparrow ;\downarrow \rangle ,|\downarrow ;\uparrow
\rangle ,|\downarrow ;\downarrow \rangle \right\} $ through the spin
independent and dependent tunnelings $t_{\sigma }$ and $\lambda $. The
couplings are different for fermions and bosons, as illustrated in Fig.~ \ref%
{fig-BF}.

\begin{figure}[t]
\centering
\includegraphics[width=3.2in]{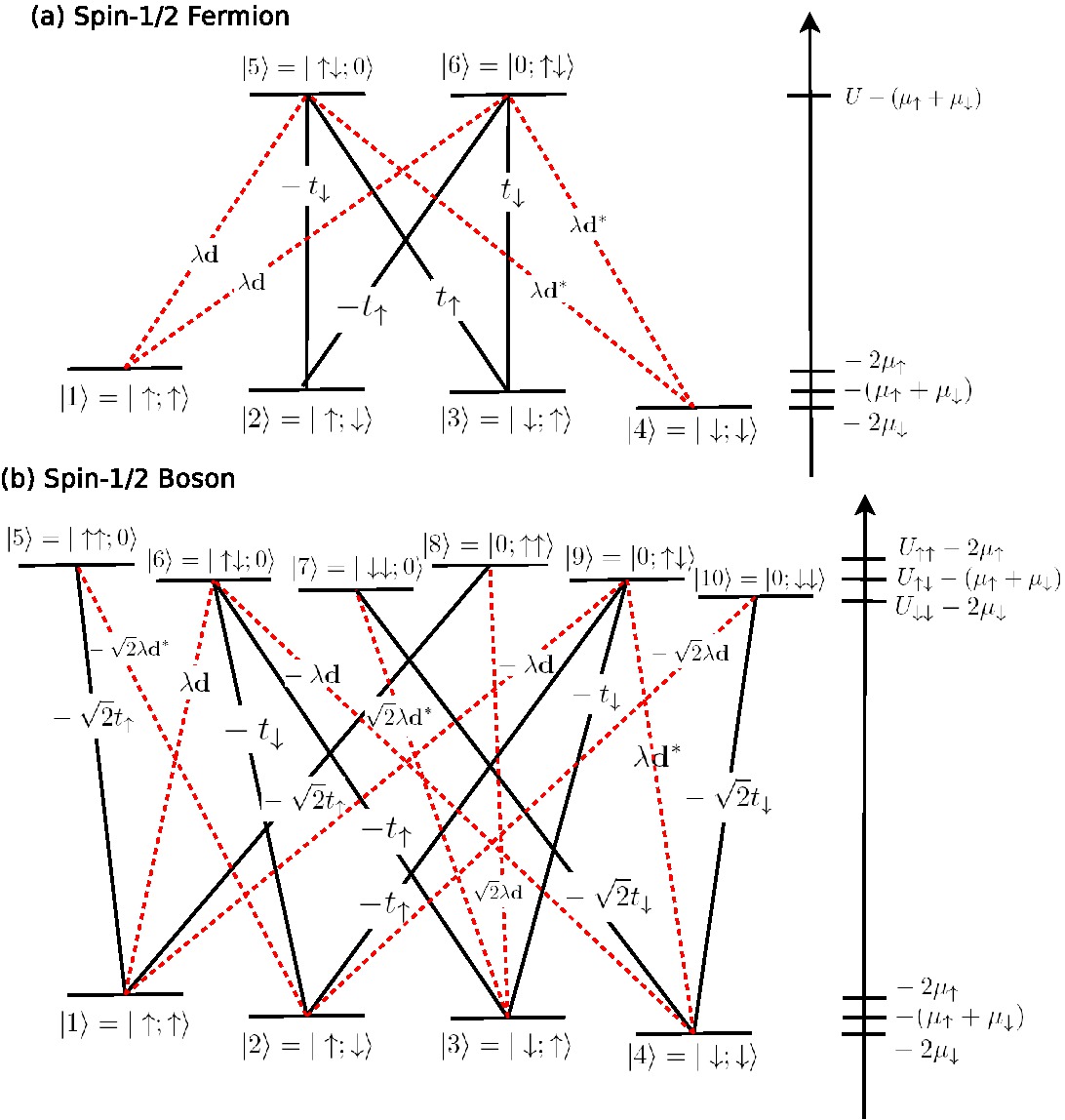}
\caption{{\bf\sffamily Transition processes due to different tunneling mechanisms.} Spin-conserving tunneling (solid lines, $t_{\protect\sigma }$ terms) and SO coupling mediated tunneling (dashed lines, $\protect\lambda $ terms) are plotted for spin-1/2 fermions (a) and spin-1/2 bosons (b). $\protect\mu_{\sigma }$ is the chemical potential. The lowest 4 levels are ground states, and the higher energy
levels are the excited states. }
\label{fig-BF}
\end{figure}

\noindent \textbf{{\small Fermionic atoms.}} For fermionic atoms, there are only two possible
excited states $|\text{ex}\rangle $ = $|\uparrow \downarrow ;0\rangle $ and $%
|0;\uparrow \downarrow \rangle $, as shown schematically in Fig.~ \ref%
{fig-BF} (a). We find $(J_{x}+J_{y})/2=4t_{\uparrow }t_{\downarrow }/U$, $%
(J_{x}-J_{y})/2=8(-dx^{2}+dy^{2})U\lambda ^{2}/(U^{2}-\Upsilon ^{2})$, and $%
J_{z}=2(t_{\uparrow }-t_{\downarrow })^{2}/U-4d^{2}U\lambda
^{2}/(U^{2}-\Upsilon ^{2})$, with $d^{2}=dx^{2}+dy^{2}$. The DM interaction
coefficient is $\mathbf{D}=2(t_{\uparrow }+t_{\downarrow })(\Upsilon
^{2}-2U^{2})\lambda /(U(\Upsilon ^{2}-U^{2}))(dy,dx,0)$, and the effective
Zeeman field contains $\mathbf{B}=4(\Upsilon -2d^{2}\Upsilon \lambda ^{2}/(\Upsilon
^{2}-U^{2}))(0,0,1)$. Note that without SO coupling the model reduces to the
well-known XXZ Heisenberg model with rotational symmetry \cite{Kuklov,Duan}.
However, this symmetry is broken by the SO coupling, yielding an XYZ-type
Heisenberg model. Similar results are also observed for bosons.

\noindent \textbf{{\small Bosonic atoms.}} For bosonic atoms, there are six excited states $|%
\text{ex}\rangle $ = $|\uparrow \uparrow ;0\rangle $, $|\uparrow \downarrow
;0\rangle $, $|\downarrow \downarrow ;0\rangle $,$|0;\uparrow \uparrow
\rangle $, $|0;\uparrow \downarrow \rangle $, $|0;\downarrow \downarrow
\rangle $, as shown in Fig.~ \ref{fig-BF} (b). Without SO coupling, the only
allowed inter-state second-order transition is between $|2\rangle $ and $%
|3\rangle $, similar to the fermionic case. The presence of SO coupling
permits other inter-state transitions, therefore the bosonic case is much
more complex than the fermionic case. For simplicity we only show the
results for $U_{\uparrow \uparrow }=U_{\downarrow \downarrow }=U_{\uparrow
\downarrow }=U$, which yields $(J_{x}+J_{y})/2=-4t_{\uparrow }t_{\downarrow
}/U$, $(J_{x}-J_{y})/2=4(d_{x}^{2}-d_{y}^{2})U\lambda ^{2}/(U^{2}-\Upsilon
^{2})$, $J_{z}=-4t_{\uparrow }t_{\downarrow }/U+2\left[ (\Upsilon
^{2}-U^{2})(t_{\uparrow }-t_{\downarrow })^{2}+2U^{2}d^{2}\lambda ^{2}\right]
/U(U^{2}-\Upsilon ^{2})$, $\mathbf{D}=-2(t_{\uparrow }+t_{\downarrow
})(\Upsilon ^{2}-2U^{2})\lambda /U(\Upsilon ^{2}-U^{2})(dy,dx,0)$, and $%
\mathbf{B}=(0,0,4\Upsilon )$.

The last term in Eq. \ref{eff-HF} reads as $\mathbf{S}_{i}\cdot \Gamma
_{ij}\cdot \mathbf{S}_{j}=\eta {\frac{8d_{x}d_{y}U\lambda ^{2}}{%
(U^{2}-\Upsilon ^{2})}}(S_{i}^{x}S_{j}^{y}+S_{j}^{y}S_{i}^{x})$, where $\eta
=+1(-1)$ for fermions (bosons). This term arises from the coupling between
states $|1\rangle $ and $|4\rangle $, $|1\rangle \langle
4|=S_{i}^{x}S_{j}^{x}-S_{i}^{y}S_{j}^{y}+i(S_{i}^{x}S_{j}^{y}+S_{i}^{y}S_{j}^{x})
$. Here the real part contributes asymmetric terms to the Heisenberg model,
while the imaginary part contributes to $\Gamma _{ij}$. In a square lattice
with $d_{x}d_{y}=0$, this term vanishes. However, for tilted lattices, such
as triangular and honeycomb, this term should be significant.

\noindent \textbf{{\small Lattice parameters.}} We estimate the possible parameters that can be
achieved in a square optical lattice $V(x,y)=V(x)V(y)$, where $%
V(x)=V_{\text{L}}\sin ^{2}(k_{L}x)$.  We define the lattice depth $s=V_{\text{L}}/E_{R}$ in units of he recoil energy $%
E_{R}=\hbar ^{2}k_{L}^{2}/2m$, where $k_{L}$
is the wavevector of the laser. The SO coupling coefficient is given by $\gamma \sim \hbar
k_{R}/m$, $k_{R}$ is the wavevector of the external Raman lasers, and $%
k_{R}\sim k_{L}$ in most cases. The Raman lasers are pure plane waves, and
serve as a perturbation to the hopping between adjacent sites.

\begin{figure}[t]
\centering
\includegraphics[width=3.3in]{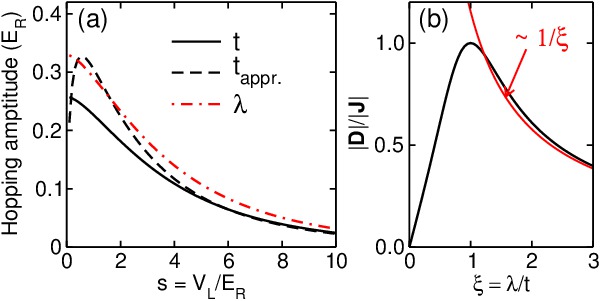}
\caption{{\bf\sffamily Tunable parameters in an optical lattice.} (a) Tunneling amplitudes as a function of lattice
depth. $t$ is the hopping due to the kinetic energy, $t_{\text{appr.}}$ is
the analytic expression derived in the deep lattice regime, and $\protect\lambda $
is the SO mediated hopping strength. (b) Plot of $|\mathbf{D}|/|\mathbf{J}| $
as a function of $\protect\lambda /t$ for $U_{\protect\sigma \protect\sigma %
^{\prime }}=U$, $t_{\protect\sigma }=t$.}
\label{fig-wannier}
\end{figure}

We use the Wannier functions of the lowest band without SO coupling to
calculate the tight binding parameters $t$ and $\lambda$. In a square lattice, coordinates decouple and the Bloch
functions are Mathieu functions. The Wannier functions can be obtained from
the Fourier transform of the Bloch functions. Our numerical results are
presented in Fig.~ \ref{fig-wannier} (a). The large $s$ limit, $t\sim t_{%
\text{appr.}}=4E_{R}/\sqrt{\pi }s^{3/4}\exp (-2\sqrt{s})$, is also plotted
for comparison. Note that $U/E_{R}\sim (8/\pi )^{1/2}k_{L}a_{s}s^{3/4}$ is in
general much larger than $t$ and can be controlled through a Feshbach
resonance independently.

In Fig.~\ref{fig-wannier} (b) we plot $|\mathbf{D}|/|\mathbf{J}|$ as a
function of $\xi =\lambda /t$ for $U_{\sigma \sigma ^{\prime }}=U$, $%
t_{\sigma }=t$. $|\mathbf{D}|/|\mathbf{J}|$ reaches the maximum value of 1.0
at $\lambda =t$. This is in sharp contrast to models of weak multiferroic
effects in solids with $D/J = |\mathbf{D}|/|\mathbf{J}| \sim 0.001-0.1$,
which is generally induced by small atomic displacements \cite{Dagotto}.
Optical lattices, by contrast, can be tuned to exhibit either weak or strong
DM terms. This enhanced tunability enables optical lattice systems to single
out the effects of strong DM interactions and study the impact of the DM
term.

There are notable differences between our model and corresponding models in
solids ($i$) In solids the SO coupling arises from intrinsic (atomic) SO
coupling and $\mathbf{D}$ is generally along the $z$ direction (out of
plane). However, in our model $\mathbf{D}$ is in the plane and the out of
plane component is zero. ($ii$) In our effective spin model, $J_{ij}^{\alpha
}$ depends on the direction of the bond ($d_{x},d_{y}$) and the SO coupling
strength, while in solids $J_{ij}^{\alpha }$ is independent of SO coupling
due to its negligible role.

\noindent \textbf{\small {Spiral order and multiferroics in 2D optical lattices.}} We now explore
the rich phase diagrams of the effective spin Hamiltonian using classical MC
simulations. Classical MC has been widely used to explore the phase
diagrams of the Heisenberg model with DM interactions in the context of
solids \cite{Liang-1,Liang-2, Liang-3, Jing} (thus weak DM interactions).
This method may not be used to determine the precise boundaries between
different phases but can be an efficient tool to determine different
possible phases. Due to the unique features of our effective model (e.g.,
strong DM interactions) the phase diagrams we present here are much more
rich and comprehensive than those explored in the context of solids. We
focus on the regime where $t_{\sigma }=t$, $U_{\sigma \sigma ^{\prime }}=U$
(spin independent), and $\Upsilon \ll U$, and define $J_{0}=4t^{2}/U$ as the
energy scale. The rescaled effective Hamiltonian becomes
\begin{equation}
H=\eta \sum_{ij}\sum_{a=x,y,z}j^{a}S_{i}^{a}S_{j}^{a}+\mathbf{D}\cdot
\mathbf{S}_{i}\times \mathbf{S}_{j}+h\sum_{i}S_{i}^{z},  \label{totalmodel}
\end{equation}%
where $j^{x}=-1+(d_{x}^{2}-d_{y}^{2})\xi ^{2}$, $%
j^{y}=-1-(d_{x}^{2}-d_{y}^{2})\xi ^{2}$, $j^{z}=-1+\xi ^{2}$, $\mathbf{D}%
=-2\xi (d_{y},d_{x},0)$, and $\xi =\lambda /t$.

\begin{figure}[b]
\centering
\includegraphics[width=3.4in]{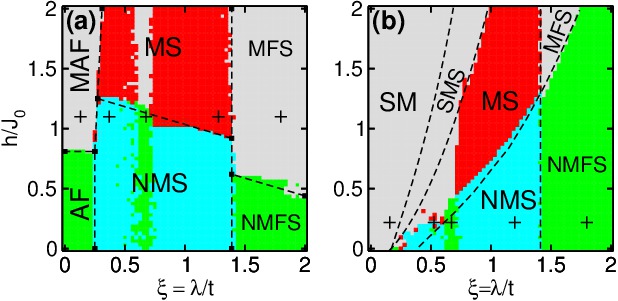}
\caption{{\bf\sffamily Phase diagrams of 2D optical lattices.} Classical Monte Carlo simulations are performed for an $8\times 8$
lattice with fermions (a) and bosons (b) at temperature $T=0.05J_{0}$. The phases diagrams are determined by
the magnetization order, the spiral order, and the spin structure factor.
Different regions correspond to: $M=0$, $P=0$ for green, $M\neq 0$, $P=0$
for grey, $M=0,P\neq 0$ for cyan, and $M\neq 0$, $P\neq 0$ for red. The
abbreviations are: (a) AF: antiferromagnetic phase with zero total
magnetization; MAF: antiferromagnetic phase with non-zero total
magnetization; NMS: zero magnetization spiral order; MS: magnetic spiral
order; NMFS: nonmagnetic flux spiral phase; MFS: magnetic flux spiral phase.
In (b), SM: simply magnetic order; SMS: simply magnetic spiral order: Other
abbreviations are the same as in (a). The dashed lines are guides to the
eye. The spin structure factors of the points marked by plus signs are shown
in Fig. \protect\ref{fig-ss}.}
\label{fig-phasediag}
\end{figure}

Eq.~\ref{totalmodel} hosts a variety of magnetic and spin spiral phases,
which are generally characterized by the magnetic and spiral order
parameters \cite{Mostovoy, Nagaosa}
\begin{equation}
M=N_{s}^{-1} \sum_{i}S_{i}^{z}\text{ and}\quad \mathbf{P}=N_{s}^{-1} \sum_{\langle i,j\rangle }%
\mathbf{d}_{ij}\times \mathbf{S}_{i}\times \mathbf{S}_{j},
\end{equation}%
where $N_{s}$ is the number of sites. However, these two order parameters do not fully characterize the phase
diagrams because in some cases there are still local magnetic or spiral
orders although both $M$ and $P=|\mathbf{P}|$ are vanishingly small. In
these cases, we also take into account the spin structure factor:
\begin{equation}
S(\mathbf{k})=N_{s}^{-2}\sum_{i,j}\langle \mathbf{S}_{i}\cdot \mathbf{S}_{j}\rangle
\exp (i\mathbf{k}\cdot (\mathbf{R}_{i}-\mathbf{R}_{j})).
\end{equation}%
$S(\mathbf{k})$ shows peaks at different positions in momentum space for different
phases. For instance, the peak of the spin structure factor is at $\mathbf{k}%
=(0,0)$ for ferromagnetic phases, $\mathbf{k}=(\pi ,\pi )$ for
antiferromagnetic phases, and $(\pi ,0)$ (or $(0,\pi )$) for the flux spiral
phase ($P=0$ but with nontrivial local spin structure). General spiral
orders correspond to other $\mathbf{k}$. We obtain the phase diagrams by
analyzing both the order parameters and spin structure factors. We have not
checked for long range order in the spin structure factor. We expect
quasi-long range order to accompany magnetized phases at low $h$, e.g., a
ferromagnetic phase for $\xi \ll 1$.

\begin{figure}[t]
\centering
\includegraphics[width=3.4in]{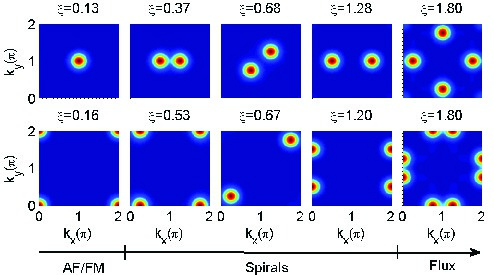}
\caption{{\bf\sffamily Spin structure factors for different quantum phases marked by plus
signs in Fig. \protect\ref{fig-phasediag}}. The upper panels show the results
for fermions at $h/J_{0}=1.1$, while the lower panels show the results for
bosons at $h/J_{0}=0.218$. }
\label{fig-ss}
\end{figure}

{The phase diagrams of an $8\times 8$ lattice in Fig. \ref{fig-phasediag} show a rich interplay between magnetic orders and spin
spiral orders. For instance, }for fermions with small SO coupling ($\xi
<0.25 $), the ground states are anti-ferromagnetic states with zero
(non-zero) magnetization for a Zeeman field $h/J_{0}<0.8$ ($h/J_{0}>0.8$).
While for large SO coupling ($\xi >1.45$), the ground states are either
nonmagnetic or magnetic flux spiral phases (similar to the flux phase with a
small spiral order $P$). For $\xi \gg 1$ the DM term is not important
because $D/J\sim 1/\xi $, therefore the pure flux phase with zero spiral
order can be observed. Similarly, the increasing SO coupling for bosonic
atoms gives rise to a series of transitions from simply magnetic
(ferromagnetic at small $h$) order to simply magnetic spiral order (with
zero total spiral order but local spiral structure), then to magnetic spiral
orders (or non-magnetic spiral orders) and finally to flux spiral orders.
The emergence of the spiral order and flux order with increasing SO coupling can be
clearly seen from the change of the spin structure factors in Fig. \ref%
{fig-ss}, which shift from $\mathbf{k}=\left( 0,0\right) $ or $(\pi ,\pi )$
to $(\pi ,0)$ and $(0,\pi )$.

\begin{figure}[tbp]
\centering
\includegraphics[width=3.1in]{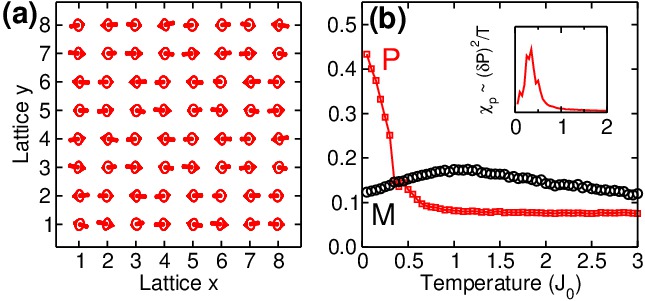}
\caption{ {\bf\sffamily Spin configurations and phase transitions.} (a) The spin configuration of fermions in an $8\times 8$ lattice at
$T=0.05J_{0}$, $\protect\xi =1.0$ and $h/J_{0}=1.5$. The corresponding
magnetization and spiral order as a function of temperature is shown in (b).  
The inset plots $\protect\chi _{p}\sim (\protect\delta P)^{2}/T$ vs. 
temperature, which indicates a phase transition at $T_{c}\sim 0.5J_{0}$.
Similar features can also been found for bosons with the same parameters.}
\label{fig-T}
\end{figure}

The spin spiral order phase transition temperature is comparable to the
magnetic phase transition temperature, $\sim J_{0}$. In Fig.~ \ref{fig-T}
(a), we plot the spin configuration of fermions at $T=0.05J_{0}$, $\xi =1.0$
and $h=1.5$ (MS phase), which shows clear spiral ordering. The corresponding
order parameters $P$ and $M$ are plotted in Fig.~ \ref{fig-T} (b) as a
function of temperature. The inset shows the susceptibility $\chi _{p}\sim
(\delta P)^{2}/T$. We see a phase transition at $T_{c}\sim 0.5J_{0}$, which
is comparable to the magnetic critical temperature \cite{Duan} (In 2D, the
Heisenberg model has a critical temperature $T_{c}=J_{0}$ in mean-field
theory). Note that spiral order can also exist in the frustrated model
without SO coupling, however, the critical temperature is generally much
smaller than the magnetic phase transition temperature \cite{Jing,Blake}.
Our results therefore show that SO coupling in the absence of frustration
provides an excellent platform to search for spiral order 
and multiferroics-based states in optical lattices.

\section*{\bf Discussion}
Finally we note that different spiral orders may be observed using optical Bragg scattering methods \cite{Corcovilos}, which probe different
spin structure factors for different spiral orders. Similar methods have
been widely used in solid state systems. Furthermore, in optical lattices,
the local spin magnetization at each lattice site (thus the magnetic order $%
M $) as well as the local spin-spin correlations (thus the spiral order $P$)
can be measured directly \cite{Greiner2,Bloch}, which provides a powerful
new tool for understanding the physics of spiral orders and multiferroic
effects in optical lattices.

\noindent  \textbf{\small {Note added.}} During the preparation of this manuscript
(the initial version is available at arXiv:1205.6211) we became aware of
work \cite{Radic,Cole,Wu} on similar topics.

\section*{\bf Methods}
The phase diagrams of an $8\times 8$ lattice are computed by classical MC methods for both fermions and bosons. The results are obtained after $10^{6}$ thermalization steps followed by $10^{6}$ sampling steps in each MC run at low temperature ($T=0.05J_{0}$). We have checked that for lower temperatures the phase diagrams do not change quantitatively. We also verify that similar phase diagrams can be obtained for larger system sizes, however, the spiral orders in a larger optical lattice become more complicated, and the boundary between different quantum phases is shifted.

{\noindent\bf{Acknowledgements}}\\
M.G. thanks S. Liang for numerical assistance with classical MC simulations. This work is supported by AFOSR
(FA9550-11-1-0313), DARPA-YFA (N66001-10-1-4025, N66001-11-1-4122), ARO (W911NF-09-1-0248), and the Jeffress Memorial Trust
(J-992). M.G. is also supported by Hong Kong RGC/GRF Projects (No. 401011, No. 401213 and No. 2130352), University Research Grant
(No. 4053072) and The Chinese University of Hong Kong (CUHK) Focused Investments Scheme. \\

{\noindent\textbf {Author contributions}}\\
M.G. and C.Z. conceived the idea, M.G., Y.Q., M.Y. performed the calculation, with input from V.W.S. and C.Z.. V.W.S. and C.Z. supervised the whole research project. All authors analyzed and discussed the results and contributed in writing the manuscript. All authors have given approval to the final version of the manuscript.\\

{\noindent\bf{Additional information}}\\
{ Competing financial Interests:} The authors declare no competing financial interests.


\begin{thebibliography}{99}

\bibitem{reviews-1} Fiebig, M. Revival of the magnetoelectric effect. \textit{J. Phys. D: Appl. Phys.} \textbf{38}, R123 (2005).
\bibitem{reviews-2} Dawber, M., Rabe, K. M. $\&$ Scott, J. F. Physics of thin-film ferroelectric oxides. \textit{Rev. Mod. Phys.} \textbf{77}, 1083 (2005).
\bibitem{reviews-3}  Basov, D. N. \emph{et al.} Electrodynamics of correlated electron materials. \textit{Rev. Mod. Phys.} \textbf{83}, 471 (2011).
\bibitem{reviews-4} Catalan, G., Seidel, J., Ramesh, R. $\&$ Scott, J. F. Domain wall nanoelectronics. \textit{Rev. Mod. Phys.} \textbf{84}, 119 (2012).
\bibitem{MF-1} Tokura, Y. $\&$ Seki, S. Multiferroics with spiral spin orders.  \textit{Adv. Mater.} \textbf{22}, 1554 (2010).

\bibitem{MF-2} Kimura, T. Spiral magnets as magnetoelectrics. \textit{Annu. Rev. Mater. Res.} \textbf{37}, 387-413 (2007).

\bibitem{MF-3} Cheong, S.-W. $\&$ Mostovoy, M. Multiferroics: a magnetic twist for ferroelectricity. \textit{Nature Materials} \textbf{6}, 13 (2007).

\bibitem{MF-4} Ramesh, R. $\&$ Spaldin, N. A. Multiferroics: progress and prospects in thin films. \textit{Nature Materials} \textbf{6}, 21 (2007).

\bibitem{MF-5} Eerenstein, W., Mathur, N. D. $\&$ Scott, J. F. Multiferroic and magnetoelectric materials. \textit{Nature} \textbf{442}, 759 (2006).

\bibitem{MF-6} Tokura, Y. Multiferroics as quantum electromagnets. \textit{Science} \textbf{312}, 1481 (2006).

\bibitem{Jing} Jin, G., Cao, K., Guo, G.-C. $\&$ He, L. Origin of ferroelectricity in high-$T_c$ magnetic ferroelectric CuO. \textit{Phys. Rev. Lett.} \textbf{108}, 187205 (2012).

\bibitem{Sergienko} Sergienko, I. A. $\&$ Dagotto, E. Role of the Dzyaloshinskii-Moriya interaction in multiferroic perovskites. \textit{Phys. Rev. B} \textbf{73}, 094434 (2006).

\bibitem{Katsura} Katsura, H., Nagoasa, N. $\&$ Balatsky, A. V. \textit{Phys. Rev. Lett.}
\textbf{95}, 057205 (2005).

\bibitem{Greiner} Greiner M., Mandel, O., Esslinger, T., H{\"a}nsch, T. W. $\&$ Bloch, I. Quantum phase transition from a superfluid to a Mott insulator in a gas of ultracold atoms. \textit{Nature} \textbf{415}, 39 (2002).

\bibitem{blochreview} Bloch, I., Dalibard, J. $\&$ Zwerger, W. Many-body physics with ultracold gases. \textit{Rev. Mod. Phys.} \textbf{80}, 885 (2008).

\bibitem{Kuklov} Kuklov, A. B. $\&$ Svistunov, B. V.  Counterflow superfluidity of two-species ultracold atoms in a commensurate optical lattice. \textit{Phys. Rev. Lett.} \textbf{90}, 100401 (2003).

\bibitem{Duan} Duan, L. -M., Demler, E. $\&$ Lukin, M. D. Controlling spin exchange interactions of ultracold atoms in optical lattices. \textit{Phys. Rev. Lett.} \textbf{91}, 090402 (2003).

\bibitem{Altman} Altman, E., Hofstetter, W., Demler, E. $\&$ Lukin, M. D. Phase diagram of two-component bosons on an optical lattice. \textit{New Journal of Physics} \textbf{5}, 113, (2003).

\bibitem{Trotzky} Trotzky, S. \emph{et al.} Time-resolved observation and control of superexchange interactions with ultracold atoms in optical lattices. \textit{Science} \textbf{319}, 295 (2008).

\bibitem{Struck} Struck, J. \emph{et al.} Quantum simulation of frustrated classical magnetism in triangular optical lattices. \textit{Science} \textbf{333}, 996 (2011).

\bibitem{NIST}  Lin, Y.-J., Jimenez-Garcia, K. $\&$ Spielman, I. B. Spin–orbit-coupled Bose–Einstein condensates. \textit{Nature} \textbf{471}, 83 (2011).

\bibitem{Pan} Chen, S. \emph{et al.} Collective dipole oscillations of a spin-orbit coupled Bose-Einstein condensate.\textit{Phys. Rev. Lett.} \textbf{109}, 115301 (2012).

\bibitem{Zhangjing1} Fu, Z., Wang, P., Chai, S., Huang, L. $\&$ Zhang, J. Bose-Einstein condensate in a light-induced vector gauge potential using 1064-nm optical-dipole-trap lasers. \textit{Phys. Rev. A} \textbf{84}, 043609 (2011).

\bibitem{Zhangjing2} Wang, P. \emph{et al.} Spin-orbit coupled degenerate Fermi gases. \textit{Phys. Rev. Lett.} \textbf{109}, 095301 (2012).

\bibitem{Ian} Jim\'{e}nez-Garc\'{\i}a, K. \emph{et al.} Peierls Substitution in an Engineered Lattice Potential. \textit{Phys. Rev. Lett.} \textbf{108}, 225303 (2012).

\bibitem{Zwierwei}  Cheuk, L. W. \emph{et al.} Spin-Injection Spectroscopy of a Spin-Orbit Coupled Fermi Gas. \textit{Phys. Rev. Lett.} \textbf{109}, 095302 (2012)

\bibitem{Yongping1} Zhang, Y. $\&$ Zhang, C. Bose-Einstein condensates in spin-orbit-coupled optical lattices: Flat bands and superfluidity. \textit{Phys. Rev. A} \textbf{87}, 023611 (2013)

\bibitem{Lin1} Lin, F., Zhang, C., $\&$ Scarola, V.W. Emergent Kinetics and Fractionalized Charge in 1D Spin-Orbit Coupled Flatband Optical Lattices \textit{Phys. Rev. Lett.} \textbf{112}, 110404 (2014).

\bibitem{Kohl} K\"{o}hl Köhl, M., Moritz, H., Stöferle, T., G\"{u}nter, K. $\&$ Esslinger, T. Fermionic atoms in a 3D optical lattice, \textit{Phys. Rev. Lett.} \textbf{94}, 080403 (2005).

\bibitem{YPPRL} Zhang, Y. $\&$  Zhang, C. Mean-field dynamics of spin-orbit coupled Bose-Einstein condensates. \textit{Phys. Rev. Lett.} \textbf{108}, 035302 (2012).

\bibitem{YP}  Zhang, Y., Chen, G. $\&$ Zhang, C. Tunable spin-orbit coupling and quantum phase transition in a trapped Bose-Einstein condensate. \textit{Sci. Rep.} \textbf{3}, 1937 (2013)

\bibitem{WS} Hewson, A. C. The Kondo Problems to Heavy Fermions. (Cambridge
University Press, Cambridge, England, 1997).

\bibitem{DM-1}  Dzyaloshinskii, I. E. Theory of helicoidal structures in antiferromagnets. I. nonmetals. \textit{Sov. Phys. JETP} \textbf{19}, 960 (1964).

\bibitem{DM-2}  Moriya, T. Anisotropic superexchange interaction and weak ferromagnetism. \textit{Phys. Rev.} \textbf{120}, 91 (1960).

\bibitem{Dagotto} Sergienko, I. A. $\&$ Dagotto, E. Role of the Dzyaloshinskii-Moriya interaction in multiferroic perovskites. \textit{Phys. Rev. B} 73, 094434 (2006).

\bibitem{Liang-1} Sen, C., Liang, S. $\&$ Dagotto, E. Complex state found in the colossal magnetoresistance regime of models for manganites. \textit{Phys. Rev. B} \textbf{85}, 174418 (2012).

\bibitem{Liang-2} Liang, S., Daghofer, M., Dong, S., Sen, C. $\&$ Dagotto, E. Emergent dimensional reduction of the spin sector in a model for narrow-band manganites. \textit{Phys. Rev. B} \textbf{84}, 024408 (2011).

\bibitem{Liang-3}  Dong, S.\emph{et al.} Exchange bias driven by the Dzyaloshinskii-Moriya interaction and ferroelectric polarization at G-type antiferromagnetic perovskite interfaces. \textit{Phys. Rev. Lett.} \textbf{103},
127201 (2009).

\bibitem{Mostovoy} Mostovoy, M. Ferroelectricity in Spiral Magnets. \textit{Phys. Rev. Lett.} \textbf{96}, 067601 (2006).

\bibitem{Nagaosa}  Katsura, H., Nagaosa, N. $\&$ Balatsky, A. Spin Current and Magnetoelectric Effect in Noncollinear Magnets. \textit{Phys. Rev. Lett.} \textbf{95}, 057205 (2005).

\bibitem{Blake} Blake, G. R.\emph{et al.} Spin structure and magnetic frustration in multiferroic RMn$_{2}$O$_{5}$ (R = Tb, Ho, Dy). \textit{Phys. Rev. B} \textbf{71}, 214402
(2005).

\bibitem{Corcovilos}  Corcovilos, T. A., Baur, S. K., Hitchcock, J. M., 
Mueller, E. J., $\&$ Hulet, R. G. Detecting antiferromagnetism of atoms in an optical lattice via optical Bragg scattering \textit{Phys. Rev. A} \textbf{81}, 013415 (2010).

\bibitem{Greiner2}  Bakr, W. S., Gillen, J. I., Peng, A., F\"{o}lling, S. $\&$ 
Greiner, M. A quantum gas microscope for detecting single atoms in a Hubbard-regime optical lattice. \textit{Nature} \textbf{462}, 74 (2009).

\bibitem{Bloch}  Weitenberg, C. \emph{et al.} Single-spin addressing in an atomic Mott insulator. \textit{Nature} \textbf{471}, 319
(2011).

\bibitem{Radic}  Radic, J., Di Ciolo, A., Sun, K. $\&$ Galitski, V. Exotic quantum spin models in spin-orbit-coupled Mott insulators
\textit{Phys. Rev. Lett.} \textbf{109}, 085303 (2012) (also available at arXiv:1205.2110).

\bibitem{Cole}  Cole, W. S., Zhang, S. Z., Paramekanti, A. $\&$  Trivedi, N.
Bose-Hubbard models with synthetic spin-orbit coupling: Mott insulators, spin textures, and superfluidity. \textit{Phys. Rev. Lett.} \textbf{109}, 085302 (2012) (also available at arXiv:1205.2319).

\bibitem{Wu} Cai, Z., Zhou, X., Wu,  C. Magnetic phases of bosons with synthetic spin-orbit coupling in optical lattices. Phys. Rev. A \textbf{85}, 061605(R) (2012) (also available at arXiv:1205.3116). \\
\end{thebibliography}
\end{document}